\documentclass[prd,twocolumn,nofootinbib,showpacs,hyperref]{revtex4-1}

\usepackage{amsfonts}
\usepackage{amsmath}
\usepackage{amssymb}
\usepackage{bm}
\usepackage{dcolumn}
\usepackage{graphicx}
\usepackage{graphics}
\usepackage[latin1]{inputenc}
\usepackage{latexsym}
\usepackage{rotating}
\usepackage[colorlinks=true]{hyperref}
\usepackage[all]{hypcap} 
\usepackage{xspace} 
\usepackage[usenames]{color}
\usepackage{mathrsfs}
\usepackage{ulem}

\definecolor {darkgreen}{rgb}{0.1,0.5,0.1}
\definecolor {orange-ish}{rgb}{0.9,0.3,0.1}
\definecolor {lavender}{rgb}{0.8,0.0,0.5}

\newcommand\be{\begin{equation}}
\newcommand\ba{\begin{eqnarray}}
\newcommand\ee{\end{equation}}
\newcommand\ea{\end{eqnarray}}
\newcommand\bw{\begin{widetext}}
\newcommand\ew{\end{widetext}}

\newcommand{\nn}{\nonumber}

\newcommand{\BH}{{\mbox{\tiny BH}}}
\newcommand{\mrm}{\mathrm}
\newcommand{\lb}{\left(}
\newcommand{\rb}{\right)}
\newcommand{\lcb}{\left\{}
\newcommand{\rcb}{\right\}}
\newcommand{\lsb}{\left[}
\newcommand{\rsb}{\right]}

\begin{document}
\title{Four-Hair Relations for Differentially Rotating \\ Neutron Stars in the Weak-Field Limit} 

\author{Joseph Bretz}
\affiliation{Department of Physics, Montana State University, Bozeman, Montana 59717, USA.}

\author{Kent Yagi}
\affiliation{Department of Physics, Montana State University, Bozeman, Montana 59717, USA.}

\author{Nicol\'as Yunes}
\affiliation{Department of Physics, Montana State University, Bozeman, Montana 59717, USA.}

\begin{abstract} 

The opportunity to study physics at supra-nuclear densities through X-ray observations of neutron stars has led to in-depth investigations of certain approximately universal relations that can remove degeneracies in pulse profile models.
%
One such set of relations determines all of the multipole moments of a neutron star just from the first three (the mass monopole, the current dipole and the mass quadrupole moment) approximately independently of the equation of state.  
%
These three-hair relations were found to hold in neutron stars that rotate rigidly, as is the case in old pulsars, but neutron stars can also rotate differentially, as is the case for proto-neutron stars and hypermassive transient remnants of binary mergers.
%
We here extend the three-hair relations to differentially rotating stars for the first time with a generic rotation law using two approximations: a weak-field scheme (an expansion in powers of the neutron star compactness) and a perturbative differential rotation scheme (an expansion about rigid rotation). 
%
These approximations allow us to analytically derive approximately universal relations that allow us to determine all of the multipole moments of a (perturbative) differentially rotating star in terms of only the first four moments.  
%
These new four-hair relations for differentially rotating neutron stars are found to be approximately independent of the equation of state to a higher degree than the three-hair relations for uniformly rotating stars.
%
Our results can be instrumental in the development of four-hair relations for rapidly differentially rotating stars in full General Relativity using numerical simulations.

\end{abstract}

\date{\today}
\maketitle

\section{Introduction}

\begin{figure*}
\begin{center}
\includegraphics[width=270pt,clip=true]{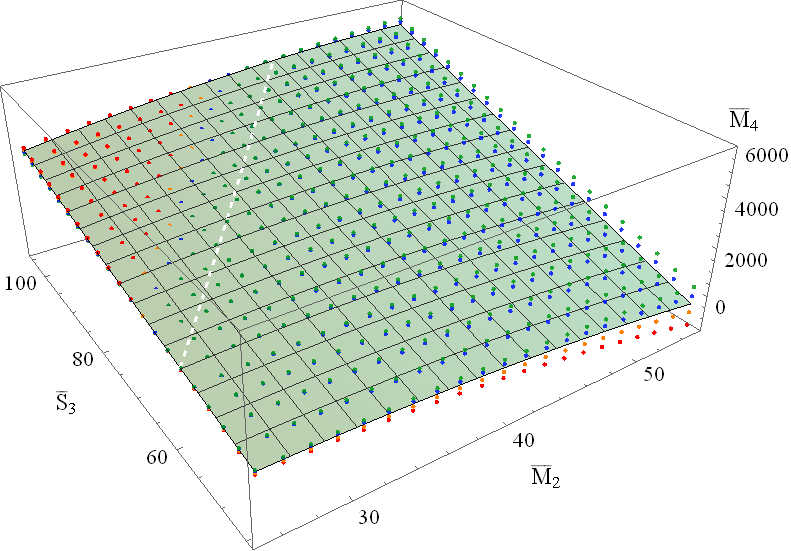}
\includegraphics[width=225pt,clip=true]{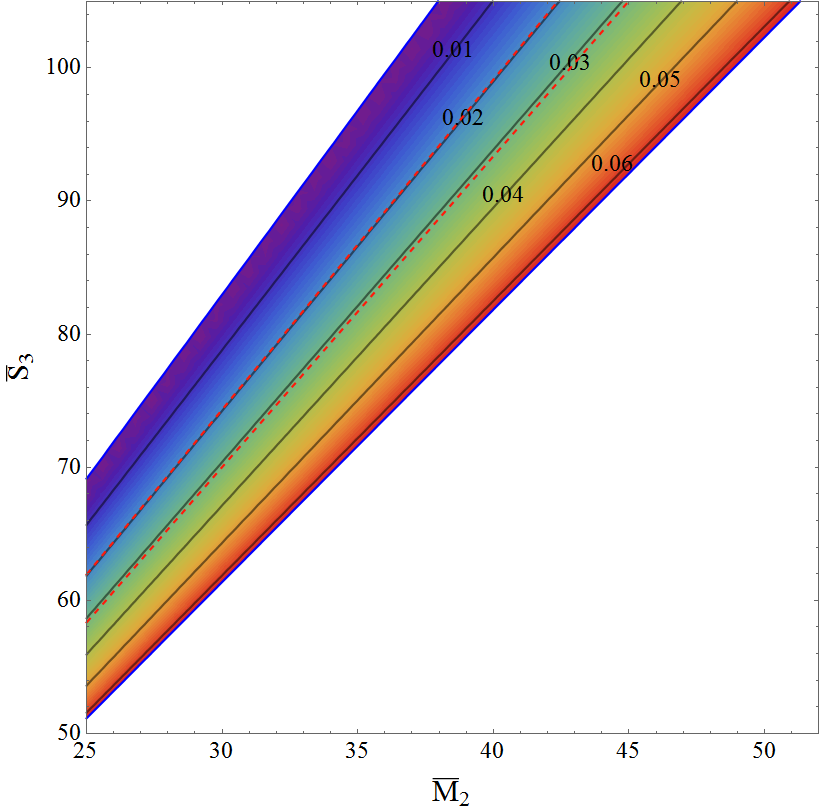}
\caption{\label{M4Plots}(color online). (Left) 3D plot of the four-hair relation showing the normalized mass hexadecapole moment $\bar{M}_4$ as a function of the normalized mass quadrupole ($\bar M_2$) and mass-current octopole ($\bar S_3$) moments computed with a polytropic EoS with various polytropic indices $n$ in the slow-rotation limit. The light blue-green plane represents the fiducial $n=0.65$ case. Observe that the $n= 0.3$ (green dots), 0.5 (blue dots), 0.8 (orange dots), 1 (red dots) results are all almost on the same plane. 
The white dashed line in the left panel represents the uniform rotation curve for an $n=1$ polytrope.
(Right) Contour and color gradient plot in the  $\bar{S}_{3}$-$\bar{M}_{2}$ plane of the maximum fractional difference between $\bar{M}_4$ computed with a polytropic EoS with index $n\in[0.3,1]$ and a fiducial index of $n=0.65$ assuming $|\gamma| < 1/2$. The red dashed lines demarcate the region of $|\gamma| < 0.1$ with $n=1$. Observe that the maximum fractional difference is always less than $\sim 6\%$ and $3\%$ for all values of $\bar{S}_{3}$ and $\bar{M}_{2}$ within the $|\gamma| < 1/2$ and $|\gamma| < 0.1$ regions respectively.
}
\end{center}
\end{figure*}

The extreme densities inside neutron stars (NSs) have sparked interest among many since new physics may lay hidden at supra-nuclear densities~\cite{L&P,Lattimer:2007nu}. One tool to probe the internal structure of NSs is electromagnetic observations of binary and isolated pulsars, although gravitational wave (GW)~\cite{read-markakis-shibata,flanagan-hinderer-love,hinderer-lackey-lang-read,lackey,damour-nagar-villain,delpozzo,Lackey:2013axa,Read:2013zra,Lackey:2014fwa,Agathos:2015uaa} and neutrino~\cite{Lattimer:2007nu} observations may soon be added to the toolbox. X-ray pulse and atomic line profiles emitted by millisecond pulsars must be fit to models that depend, in part, on intrinsic NS parameters, such as their mass, radius, moment of inertia and quadrupole moment~\cite{Morsink:2007,Lo:2013ava,Baubock:2013ato,Psaltis:2014pul}. Since the NS equation of state (EoS) -- the relation between state variables such as pressure and density -- is unknown, one must fit all of these parameters to the data \emph{independently}, thus diluting the information that can be extracted. Reducing the number of independent parameters in an EoS independent way would help extract more information from observables~\cite{Baubock:2013hta,Psaltis:2014xry}, and in particular, it may allow for the precise extraction of the NS mass and radius with NICER~\cite{2012SPIE.8443E..13G} and LOFT~\cite{2012AAS...21924906R,2012SPIE.8443E..2DF,2014SPIE.9144E..2TF}. 

One can partially break the degeneracies among parameters in NS observations by using universal relations between certain observables that do not depend strongly on the EoS. For example, the moment of inertia (I), the tidal Love number and the quadrupole moment (Q or $M_2$) have recently been found to be interrelated in an approximately EoS-independent fashion~\cite{YYb,YYa}. The I-Love-Q relations can thus be used to reduce the number of model parameters without knowledge of the high-density EoS. Such relations have already been extended to binary systems with dynamical tides~\cite{Maselli:2013}, proto-neutron stars (PNSs)~\cite{Martinon:2014uua}, magnetized NSs~\cite{Haskell:2014}, NSs with different EoSs~\cite{Lattimer:2013}, rapidly rotating NSs~\cite{Doneva:2013rha,Pappas:2014,Chakrabarti:2014,Yagi:2014rel,Stein:2014}, NSs in a post-Minkowskian expansion~\cite{Chan:2015}, NSs with anisotropic pressure~\cite{Yagi:2015ani}, as well as relations with different normalizations~\cite{Majumder:2015kfa}, for exotic compact objects~\cite{Pani:2015tga} and in modified gravity theories~\cite{YYb,YYa,Sham:2014,Doneva:2014,Pani:2014,Kleihaus:2014,Doneva:2015hsa}.

The I-Love-Q discovery has also led to the NS three-hair relations, which resemble the well-known, black hole no-hair relations~\cite{israel,israel2,hawking-uniqueness,carter-uniqueness,Hansen:1974zz,robinson}. By using a multipole expansion to describe the mass distribution inside a star, its exterior gravitational field becomes a function of the mass and mass-current multipole moments. Although the expansion involves an infinite sum of moments, the three-hair relations reduce all of the moments to functions of only the first three in a way that is approximately independent of the EoS. Therefore, the NS three-hair relations allow one to approximately describe the exterior gravitational field of a NS in terms only of its first three, non-vanishing multipole moments.   

Such approximate universality is highly sensitive to the elliptical isodensity approximation used to derive the three-hair relations~\cite{Yagi:2014qua}. This approximation assumes that constant density contours of NSs are self-similar ellipsoids, with the ellipsoidal radius identical to the radius of a spherically symmetric star of the same volume. The three-hair relations were first derived for weakly-gravitating stars~\cite{Stein:2014,Chatziioannou:2014}, and were soon after confirmed in full General Relativity (GR) up to hexadecapole order~\cite{Yagi:2014rel}, but always assuming uniform (rigid) rotation, i.e.~assuming a constant angular velocity of rotation. This assumption is well-justified for old and cold NSs, such as millisecond pulsars in which differential rotation has damped out, and it ensures the elliptical isodensity approximation holds to a sufficiently good accuracy. 

But there are other physical scenarios in which uniform rotation is not a good assumption. 
One such scenario are PNSs born after a supernova explosion, which may eventually be probed via neutrino and electromagnetic observations. Universal relations in PNSs were studied in~\cite{Martinon:2014uua}, where the original I-Love-Q relations were found to hold after only a few hundred milliseconds from birth. Such an analysis, however, was limited to uniformly rotating PNSs. 
Another scenario in which uniform rotation is not a good assumption is a hypermassive NS (HMNSs) produced after the merger of a NS binary, a prime candidate for GW detection~\cite{Shapiro:2000zh,Paschalidis:2012ff,Hotokezaka:2013iia}. Differential rotation may here be key by providing a temporary mechanism that prevents the  HMNS from further collapsing.
A third scenario in which differential rotation may arise is in NS r-modes -- toroidal oscillations with a Coriolis restoring force -- that may be driven unstable via GW emission, thus leaving an imprint in GW signals~\cite{Haskell:2012}. Stars with differential rotation have also been found to be more susceptible to non-axisymmetric instabilities, which can also lead to GW signals~\cite{Fujisawa:2015nla,Stavridis:2007xz}.

How does differential rotation affect the three-hair relations? This is the topic of this paper. In order to understand physically the effect of differential rotation in the three-hair relations, we work \emph{semi-analytically} by making use of two approximations: a weak-field scheme and a perturbative differential rotation scheme. The former is simply an expansion in the NS compactness (the ratio of its mass to its radius), while the latter is an expansion about rigid rotation. For sufficiently small deformations, these approximations allow us to continue to use the elliptical isodensity approximation; this, in turn, allows us to develop an analytic understanding of how differential rotation impacts the three-hair relations, as well as to develop new, four-hair relations that are as EoS universal as the three-hair relations for rigidly rotating stars. 

Several physical scenarios exist in which the small differential rotation approximation is a good approximation to Nature, such as during a certain phase after the birth of PNSs. When a PNS forms after gravitational collapse, it is hot, unstable and highly differentially rotating. After tens to a few hundreds of milliseconds, the PNS goes into a ``quasi-stationary'' phase~\cite{Martinon:2014uua}, where it is quieter, evolves more slowly and can be described as a sequence of equilibrium configurations~\cite{Pons:1998mm,Pons:2000xf,Pons:2001ar,Fischer:2009af}. Such a phase lasts for roughly one minute, until finally, the star becomes cold and uniformly rotating. During this evolution, the amount of differential rotation is reduced due to dissipative mechanisms, such as magnetic braking and viscous damping~\cite{Shapiro:2000zh,Cook:2003ku,Liu:2003ay}. Therefore, as a PNS evolves into a cold NS, an intermediate phase exists (the quasi-stationary phase), where the small differential rotation approximation is valid. This may not be the case for differentially rotating HMNSs formed after binary mergers, which can last $\sim 20$ ms or longer depending on the EoS, because these can collapse to a BH before reaching a phase where the small differential rotation approximation is valid.

Even within the small differential rotation approximation we must still make a choice that determines precisely how the star rotates differentially: we need to specify a \emph{rotation law}. We here work with a generic analytic parametrization~\cite{Galeazzi:2011nn} that allows us to model j-constant and v-constant laws, the Keplerian angular velocity profile and the rotation law for HMNSs. We further introduce a dimensionless parameter $\gamma$ that quantifies the extent to which the rotation law differs from a constant, with $\gamma=0$ corresponding to uniform rotation. The rotation law affects directly only the mass-current moments $S_{\ell}$, because to leading-order in the weak-field expansion the mass moments $M_{\ell}$  do not depend explicitly on rotation. Differential rotation, however, does affect the stellar shape, which does also modify $M_{\ell}$.

With this at hand, we rederive the three-hair relations of~\cite{Stein:2014} using a polytropic EoS with polytropic index $n$. We first derive the modified equations of structure analytically and then solve them numerically for $n \in [0.3,1]$ and analytically for $n=0$ and for $n = \epsilon \ll 1$ in the slow-rotation limit, extending Ref.~\cite{Chatziioannou:2014} to slowly, differentially rotating stars. We then compute various multipole moments and their interrelations and find that the three-hair relations acquire a correction that is proportional to $\gamma$. This correction is mildly EoS dependent, deteriorating the universality of the three-hair relations by at most ${\cal{O}}(\gamma \times 10\%)$. But since $|\gamma| \ll 1$, this deterioration is always much less than $10\%$, and thus, much smaller than the inherent EoS variability of the three-hair relations in the rigidly rotating case. 

These semi-analytic results suggest the construction of new \emph{four-hair} relations for differentially rotating NSs by replacing the $\gamma$ dependence of the three-hair relations with the next, independent multipole moment (the mass-current octopole $S_{3}$). The resulting four-hair relations are then independent of $\gamma$ and EoS-independent to approximately $\mathcal{O}(10\%)$, just as in the three-hair case of rigidly rotating stars. The left panel of Fig.~\ref{M4Plots} presents an example of the four-hair relations, where we plot $\bar M_4$ in terms of $\bar M_2$ and $\bar S_3$, normalized to their corresponding black hole expressions and using a polytropic EoS with $n \in [0.3,1]$. Observe that regardless the value of $n$, the multipole moments lie approximately on the same so-called invariant, four-hair plane. The right panel of Fig.~\ref{M4Plots} shows the maximum fractional difference in the $\bar{S}_{3}$-$\bar{M}_{2}$ plane between $\bar{M}_{4}$ computed with a polytropic EoS with $n \in [0.3,1]$ and a fiducial EoS with $n = 0.65$, assuming $|\gamma| < 1/2$. Observe that the maximum fractional difference stays always below $\sim 6\%$. 

How does the EoS variation in the four-hair relations compare to the three-hair relations for uniformly rotating stars? The $|\gamma| < 0.1$ region of the right panel of Fig.~\ref{M4Plots} shows that the maximum EoS variation in the four-hair relations is $\sim 3\%$, which is actually \emph{smaller} than the $\sim 4\%$ variation of the three-hair relations for uniformly rotating stars~\cite{Stein:2014}. This means that small amounts of differential rotation actually \emph{improve} the universality, albeit by a small amount. The reason for this improvement is that the four-hair relations contain an additional degree of freedom ($\bar{S}_{3}$) relative to the three-hair relations. The EoS variation can then be explored along a new direction (with a fixed $\bar M_{2}$ and $\bar S_3$) that is not allowed in the three hair case and that can be exploited to minimize the EoS variation further. Of course, for large differential rotation, i.e.~$|\gamma| > 0.1$, the EoS variation increases, and eventually, the EoS variation in the four-hair relation exceeds that in the three-hair relations with uniform rotation.

The remainder of this paper presents the computational details that lead to the above results. 
Section~\ref{sec:mult-mom} introduces the generalized differential rotation law we use in this paper and the concept of multipole moments for NSs. 
Section~\ref{sec:unv-rel} derives the three- and four-hair relations of differentially rotating stars and discusses their EoS-dependence. 
Section~\ref{sec:disc} concludes and points to future research. 
Henceforth, we follow the notation of~\cite{Stein:2014}, and in particular, we use geometric units in which $G=1=c$.

\section{Multipole Moments for Differentially Rotating Stars}
\label{sec:mult-mom}

The gravitational potential around a source, assuming an asymptotically flat and axisymmetric spacetime, can be completely described by its mass and mass-current moments. To leading- (so-called \emph{Newtonian}) order in a weak-field expansion ($C := M_{*}/R_{*} \ll 1$, with $M_{*}$ the mass and $R_{*}$ the equatorial radius), the multipole moments of differentially rotating stars are given by~\cite{Ryan}
\ba
\label{eq:Ml-integral}
M_{\ell} &=& 2 \pi  \int^{1}_{-1} \,\, \int^{R_{*}(\mu)}_0 \! \! \rho(r,\mu)  r^{\ell + 2}  dr  \,\, P_{\ell}(\mu) d\mu\,, \\
 S_{\ell} &=& \frac{4 \pi}{\ell+1}  \int^{1}_{-1} \,\, \int^{R_{*}(\mu)}_0 \Omega(r,\mu) \rho(r,\mu) r^{\ell + 3} dr \nn \\
&\times& \frac{ d P_{\ell}(\mu)}{d \mu}(1-\mu^2) d\mu\,, 
\label{eq:Sl-integral}
\ea
where $\mu=\cos{\theta}$, $R_{*}(\mu)$ is the stellar surface profile, $\rho(r,\mu)$ is the star's density, $P_\ell(\mu)$ are Legendre polynomials, and $\Omega(r,\mu)$ is the angular velocity of the star. Clearly, when $\Omega(r,\mu) = \Omega_{c} = {\rm{const.}}$, the star rotates uniformly. In the weak-field limit, the gravitational potential of axially symmetric stars can be completely described by the mass multipole moments. In general, reflection symmetry about the stellar equator forces half of the moments to vanish: $M_{2\ell +1}=0=S_{2\ell}$.

Differential rotation modifies the multipole moments in two different ways. First, the radial and polar angle dependence of $\Omega$ directly modifies the mass-current moments. This dependence is encoded in the so-called \emph{rotation-law}, for which we here adopt the generalized model of~\cite{Galeazzi:2011nn}:
\be
\label{eq:Omega}
\frac{\Omega }{\Omega _c}=\lb1-\alpha\gamma \frac{r^{2}}{{a_1}^{2}} \sin^{2}{\theta}\rb^{\frac{1}{\alpha}}\,,
\ee
where $\Omega _c$ is the angular velocity at the core and $a_1$ is the semi-major axis of the uniform rotation configuration. The parameter $\alpha$ determines the type of rotation law. For the j-constant and v-constant laws, the Keplerian angular velocity profile and the rotation law for HMNSs, $\alpha=-1,-2,-{4}/{3}$ and $-4$ respectively~\cite{Galeazzi:2011nn}. The dimensionless parameter $\gamma$ controls the amount of differential rotation, with uniform rotation recovered when $\gamma=0$, and with $\alpha$ and $\gamma$ satisfying $\alpha \gamma <0$~\cite{Galeazzi:2011nn}. The perturbative differential rotation approximation consists of assuming that $|\gamma| \ll 1$ and expanding all equations in this quantity. Such an expansion is a perturbation about the uniform rotation case, reducing the rotation law to
\be
\label{eq:OmegaExpanded}
\frac{\Omega }{\Omega _c}=1-\gamma \frac{r^{2}}{{a_1}^{2}} \sin^{2}{\theta}+\mathcal{O}(\gamma^2)\,.
\ee
Notice that the $\alpha$ has canceled out.

The second modification that differential rotation introduces is a change in the stellar shape $R_*(\mu)$. Following~\cite{Stein:2014}, we assume that $R_*(\mu)$ corresponds to the radial profile of a constant density star. In the uniformly rotating case, this corresponds to spheroids, but when when differential rotation is present the shape is no longer spheroidal. The change in stellar shape can be expressed as an infinite sum of even Legendre polynomials $P_\ell(\cdot)$: 
\ba \label{eq:DevSpher}
R_*(\mu) &=& a_1 \left(\sqrt{1+ \frac{e^2}{1-e^2} \mu^2} +  \gamma \sum _{j=0}^{\infty } P_{2j}(\mu)\alpha_{2j}(e^2) \right) \nn \\
& + & \mathcal{O}(\gamma^2)\,,
\ea
where $\alpha_{2 j}(e^2)$ is a function of the stellar eccentricity squared, $e^2 \equiv 1 - a_3^2/a_1^2$, with $a_3$ the semi-minor axis of the uniformly rotating configuration. Appendix~\ref{app:shape} presents a mathematical derivation of this result. 

\begin{figure}[h]
\begin{center}
\includegraphics[width=6.5cm,clip=true]{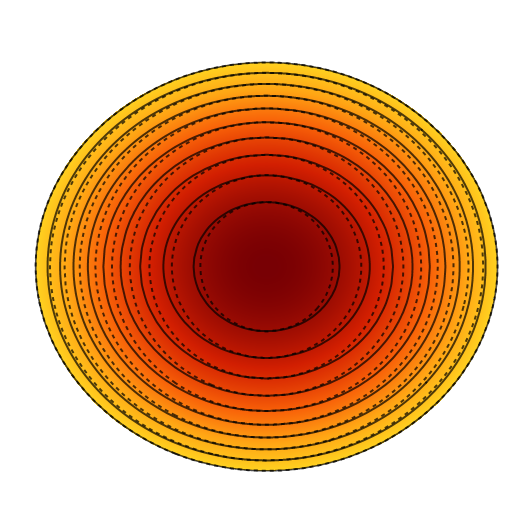}
\includegraphics[width=0.9cm,clip=true]{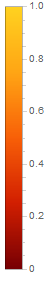}
\caption{\label{Isodense}(color online). Stellar density (color gradient) and isodensity layers under uniform rotation in the elliptical isodensity approximation (solid contours) and for a typical NS (dashed contours) constructed with a realistic SLy EoS~\cite{Douchin:2001}. The color gradient legend corresponds to the ratio of elliptical radius $\tilde{r}$ to the stellar surface $R_*$. Observe how the solid contours are close to the dashed contours, except close to the stellar core.}
\end{center}
\end{figure}

Further assumptions and approximations allow us to make more analytical progress. Following~\cite{Lai}, we adopt the elliptical isodensity approximation, where we assume the layers of constant density are ellipsoidal with the same eccentricity throughout the star. In reality, isodensity layers become more spherical towards the center, as illustrated in Figure~\ref{Isodense}, but this has been shown to have a very small effect in the structure of uniformly rotating NSs, i.e.~calculations that employ the elliptical isodensity approximation reproduce full numerical calculations up to the mass-shedding limit with errors of at most 3$\%$~\cite{Lai}. The elliptical isodensity approximation introduces an even smaller error when computing multipole moments, since the mass near the center of a star contributes less than the mass at larger radii~\cite{YYa}.

Although the elliptical isodensity approximation has only been shown to be accurate for uniformly rotating stars, one can still impose such an approximation for differentially rotating stars as long as the amount of differential rotation is small. In order to explain this point further, let us introduce a dimensionless parameter $\epsilon \ (\ll 1)$ that characterizes the difference between the three-hair relations among multipole moments for uniformly rotating stars with and without imposing the approximation. Then, expanding the three-hair relations for differentially rotating stars around both $\gamma = 0$ and $\epsilon = 0$, one can decompose the three-hair relations order by order, obtaining a bivariate series with structure $\mathcal{O}(\gamma^0 \epsilon^0)+\mathcal{O}(\gamma^0 \epsilon^1)+\mathcal{O}(\gamma^1 \epsilon^0)+\mathcal{O}(\gamma^1 \epsilon^1)+\mathcal{O}(\gamma^2, \epsilon^2)$. The first term is the three-hair relations for uniformly rotating stars within the elliptical isodensity approximation. The second term is the correction to this due to the breakage of the approximation, which has been shown to be small~\cite{Yagi:2014rel}. The third term is the three-hair relations for differentially rotating stars within the elliptical isodensity approximation, which we calculate in this paper. The fourth term is the correction to the latter due to the breakage of the approximation. But notice that this fourth term is of higher order than $\mathcal{O}(\gamma^1 \epsilon^0)$ by $\mathcal{O}(\epsilon)$. Therefore, we can safely impose the elliptical isodensity approximation even for differentially rotating stars as long as differential rotation is small and we perform all calculations perturbatively.

The elliptical isodensity approximation suggests that to solve Eqs.~\eqref{eq:Ml-integral} and~\eqref{eq:Sl-integral} we should introduce an adapted coordinate system $x^i=\tilde{r} \, \Theta(\cos\theta) \, n^i$ where $n^i=(\sin\theta \cos\phi,\sin\theta \sin\phi,\cos\theta)$ is the unit direction vector in spherical coordinates. The change in stellar shape manifests as a correction to $\Theta(\mu)$, namely
\be \label{eq:Theta}
\Theta (\mu)\equiv{\sqrt{1+ \frac{e^2}{1-e^2} \mu^2}} + \gamma \sum _{j=0}^{\infty } P_{2j}(\mu)\alpha_{2j}(e^2) +  \mathcal{O}(\gamma^2)\,,
\ee
where we used Eq.~\eqref{eq:DevSpher} and the stellar surface is defined at $\tilde{r}=a_1$. 

Changing coordinates allows the multipole moment integrals to be separated into radial and angular parts. The mass multipole moments, $M_{\ell}$, become
\be\label{eq:M_l}
M_{\ell} = 2 \pi R_{\ell}\lsb I_{\ell,3}+\gamma(\ell+3)\tilde{I}_{\ell,2}\rsb +  \mathcal{O}(\gamma^2)\,,
\ee
where~\cite{Stein:2014}
\be
\label{eq:Int}
R_\ell \equiv \int_0^{a_1} \rho(\tilde{r}) \tilde{r}^{\ell+2} d\tilde{r}\,, \qquad I_{\ell,k}\equiv\int_{-1}^{1} P_{\ell}(\mu) \Theta(\mu)^{\ell+k} d\mu\,.
\ee
Notice the density depends only on $\tilde{r}$, and the deviation from spheroidicity is encoded in $\tilde{I}_{\ell,2}$ as
\be
\label{eq:Intt}
\tilde{I}_{\ell,k}\equiv \int^1_{-1} P_{\ell}(\mu) \Theta(\mu)^{\ell+k} \sum_{j=0}^\infty P_{2j}(\mu) \alpha_{2j}(e^2)d\mu\,.
\ee
Using Eq.~\eqref{eq:OmegaExpanded} to model differential rotation, the mass-current moments, $S_{\ell}$, become
\ba
S_\ell&=&\frac{4\pi\ell}{2\ell+1} \Omega_c R_{\ell+1} \lcb\delta I_{\ell,3}+\gamma\lsb(\ell+4)\delta\tilde{I}_{\ell,2}
\right.\right.\nn\\
&-&\left.\left.
\frac{\ell(\ell+1)}{a_1^2(2\ell+1)}\lb\delta I_{\ell,5}-\delta\bar{I}_{\ell,5}\rb\rsb\rcb\ +  \mathcal{O}(\gamma^2)\,,
\ea
where $\delta N_{\ell,k} := N_{\ell-1,k+2}-N_{\ell+1,k}$ for $N = I$, $\tilde{I}$, and $\bar{I}$ with the last defined via
\be
\bar{I}_{\ell,k}\equiv\int_{-1}^{1} \mu^2 P_{\ell}(\mu) \Theta(\mu)^{\ell+k} d\mu\,.
\ee
The closed-form solutions to $I_{\ell,k}$ and $\bar{I}_{\ell,k}$ for various $k$ values can be derived using Eq.~$(7.226.1)$ from~\cite{Gradshteyn}; from these, we find
\ba
\delta I_{\ell,3}&=&(-1)^{\frac{\ell-1}{2}}\frac{  2(2\ell+1)}{\ell (\ell+2)}\sqrt{1-e^2}\;e^{\ell-1}\,,\\
\delta I_{\ell,5}&=&(-1)^{\frac{\ell+1}{2}}\frac{2   (2 \ell+1) }{\ell (\ell+2) (\ell+4)} \sqrt{1-e^2}\;e^{\ell-1}\nn\\
&\times& \lsb\lb e^2-2\rb \ell-3\rsb\,,\\
\delta\bar{I}_{\ell,5}&=&(-1)^{\frac{\ell+3}{2}}\frac{2(2\ell+1)}{\ell (\ell+2) (\ell+4)} \lb1-e^2\rb^{3/2}e^{\ell-3}\nn\\
&\times& \lsb\lb e^2-1\rb \ell+1\rsb\,,
\ea
while $\delta \tilde{I}_{\ell,k}$ will be solved later under the slow-rotation approximation.

By separating the integrals, all of the EoS-dependence that remains is in the radial integral, $R_\ell$. Let us then compute these quantities using a polytropic EoS, $p=K \rho^{1+1/n}$, where $p$ is pressure and $\rho$ is density; future work could easily extend these results to more complicated EoSs~\cite{Lattimer:2007nu,Read:2009}. We first transform to dimensionless variables, defined via $\rho = \rho_c [\vartheta(\xi)]^n$, where $\xi = (\xi_1/a_1)\tilde{r}$ is a dimensionless radius, such that $\xi=\xi_1$ corresponds to the stellar surface, and $\rho_c$ is the central density. After this transformation, the radial integral becomes $R_\ell=\rho_c (a_1/\xi_1)^{\ell+3} \mathcal{R}_{n,\ell}$, where~\cite{Stein:2014}
\be
\mathcal{R}_{n,\ell}\equiv\int_{0}^{\xi_1} [\vartheta_\mrm{sph}(\xi)]^n \xi^{\ell+2} d\xi\,.
\ee
The spherically-symmetric, Lane-Emden function $\vartheta_\mrm{sph}$ has replaced $\vartheta$ under the assumptions of the elliptical isodensity approximation, i.e.~the constant $\tilde{r}$ surfaces are assumed to have the same density profile as a spherically symmetric star of the same mass. We drop the subscript ``sph'' below for simplicity.

Now that the radial integrals have been normalized for a polytropic EoS, the multipole moments become
\ba
\label{eq:M-fin}
M_\ell &=&M_0\frac{a_1^{\ell}}{\xi _1^{\ell+2}}\frac{I_{\ell,3}}{I_{0,3}}\frac{  \mathcal{R}_{n,\ell}}{|\vartheta^\prime(\xi_1)|}\left[1+\gamma  \left((\ell+3)\frac{
   \tilde{I}_{\ell,2}}{I_{\ell,3}}-3\frac{ \tilde{I}_{0,2}}{I_{0,3}}\right)\right] \nn \\
   & + & \mathcal{O}(\gamma^2)\,,
\ea
\ba
\label{eq:S-fin}
S_\ell&=&\frac{2\ell M_0\Omega _c}{(2\ell+1)}\frac{a_1^{\ell+1}}{\xi _1^{\ell+3}}\frac{\delta I_{\ell,3}}{I_{0,3}}\frac{   \mathcal{R}_{n,\ell+1}}{ |\vartheta^\prime(\xi_1)| } \left\{1+\gamma\left[(\ell+4)\frac{\delta\tilde{ I}_{\ell,2}}{\delta I_{\ell,3}}\right.\right.\nn \\
&-& \left.\left. 3\frac{\tilde{I}_{0,2}}{I_{0,3}}- \frac{\ell(\ell+1)}{(2\ell+1)}\frac{ \left(\delta I_{\ell,5}- \delta \bar{I}_{\ell,5}\right) 
}{ \xi _1^2 \, \delta I_{\ell,3} } \frac{\mathcal{R}_{n,\ell+3} }{\mathcal{R}_{n,\ell+1}}\right]\right\} +  \mathcal{O}(\gamma^2)\,, \nn \\
\ea
where one can solve for $I_{\ell,3}$ to find 
\ba
I_{\ell,3}&=&(-1)^{\frac{\ell}{2}}\frac{2   }{\ell+1}\sqrt{1-e^2}\;e^\ell\,.
\ea
In deriving these expressions we have used Eq.~\eqref{eq:M_l} with $\ell=0$ to solve for $\rho_{c}$ in terms of the mass monopole $M_{0}$, where $|\vartheta^\prime(\xi_1)|$ is the derivative of a dimensionless function related to density at the stellar surface. Notice the differential rotation term for the mass moments does not depend on the EoS, unlike the mass-current moments.

\section{Universal Relations for Differentially Rotating Stars}
\label{sec:unv-rel}

\subsection{Three-Hair and Four-Hair Relations}

Let us begin deriving the universal three- and four-hair relations by normalizing the multipole moments, using the same normalization of~\cite{Stein:2014}:
\be
\bar{M}_{\ell}\equiv(-)^{\frac{\ell}{2}}\frac{ M_{\ell}}{M^{\ell +1} \chi ^{\ell}}\,, \qquad \bar{S}_{\ell}\equiv(-)^\frac{\ell-1}{2}\frac{ S_{\ell}}{M^{\ell+1} \chi ^{\ell}}\,,
\ee
where $\chi\equiv S_1/M_0^2$. It should be noted that $\bar{M}_0=\bar{S}_1=1$ since $M_0$ and $S_1$ are used to normalize the moments. The normalized moments $\bar{M}_{2\ell+2}$ and $\bar{S}_{2\ell+1}$ are given explicitly in Eqs. (\ref{eq:M2lp2}) and (\ref{eq:S2lp1}) of Appendix~\ref{app:full} respectively. Inspecting these equations, one notices that all of the EoS-dependence in the differential rotation corrections can be absorbed into a coefficient, 
\be
\label{eq:C-def}
\bar{C}_{n,\ell} := \frac{{\mathcal{R}_{n,2 \ell+4}}}{\xi _1^2 \, {\mathcal{R}_{n,2 \ell+2}}}\,.
\ee
Therefore, by studying how much $\bar{C}_{n,\ell}$ varies with $n$ for a set of $\ell$, we can determine how much that mass and mass-current multipole moments vary with the EoS. 

Deriving the \textit{first universal relation} begins by solving $\bar{M}_{2}$ for $a_1^2$ and substituting this into $\bar{M}_{2\ell+2}/\bar{S}_{2\ell+1}$. Retaining only terms up to ${\cal{O}}(\gamma)$, one finds  
\be \label{eq:FirstUniv}
\bar{M}_{2\ell+2}=\bar{M}_2\bar{S}_{2\ell+1}\lb1 + \gamma\,\alpha_{n,\ell}^{(1)}\rb +  \mathcal{O}(\gamma^2)\,,
\ee
where the differential rotation term $\alpha_{n,\ell}^{(1)}$ depends on $\bar{C}_{n,\ell}$ and can be found in Eq.~\eqref{eq:alpha1}. Notice that Eq.~\eqref{eq:FirstUniv} agrees with the uniform rotation result of~\cite{Stein:2014} when $\gamma=0$.

Deriving the \textit{second universal relation} begins by solving $\bar{S}_{2\ell+1}$ for $a_1^{2\ell}$ and inserting $a_1^{2\ell}(\bar{S}_{2\ell+1})$ into $\bar{M}_{2\ell+2}$. Retaining terms up to $\mathcal{O}(\gamma)$, one finds
\be \label{eq:SecondUniv}
\bar{M}_{2\ell+2}=\bar{A}_{n,\ell}(\bar{S}_{2\ell+1})^{1+1/\ell}\lb1+\gamma\,\alpha_{n,\ell}^{(2)}\rb+  \mathcal{O}(\gamma^2)\,,
\ee
where a second coefficient defined in~\cite{Stein:2014}
\be
\bar{A}_{n,\ell}\equiv\frac{(2 \ell+3)^{1/\ell} \left(\mathcal{R}_{n,2}^{1+1/\ell} \mathcal{R}_{n,2 \ell+2}^{-1/\ell}\right)}{3^{1+1/\ell} \xi_1^2 |\vartheta^\prime(\xi_1)|}
\ee
absorbs the EoS-dependence. The differential rotation term $\alpha_{n,\ell}^{(2)}$ can be found in Eq.~(\ref{eq:alpha2}).

By combining these universal relations, all of the multipole moments are related through the \textit{three-hair relations}
\begin{align} 
\label{eq:3HR}
\bar{M}_{2\ell+2}+i\bar{S}_{2\ell+1} &= 
\bar{B}_{n,\ell}\bar{M}_2^{\ell} 
\left[ \bar{M}_2 \left(1+\gamma\,\alpha_{\ell}^{(3,\bar{M})} \right)
\right. 
\nn \\
& \left.
+ i \, \bar{S}_1 \left(1+\gamma\,\alpha_{n,\ell}^{(3,\bar{S})} \right)
\right] +  \mathcal{O}(\gamma^2)\,,
\end{align}
where a third coefficient, also from~\cite{Stein:2014},
\be
\bar{B}_{n,\ell}\equiv\frac{3^{\ell+1} \xi _1^{2\ell} |\vartheta^\prime(\xi_1)|^\ell  }{2\ell+3}
\frac{\mathcal{R}_{n,2\ell+2}}{\mathcal{R}_{n,2}^{\ell+1}}
\ee
again absorbs the EoS-dependence. The differential rotation corrections $\alpha_{\ell}^{(3,\bar{M})}$ and $\alpha_{n,\ell}^{(3,\bar{S})}$ are given in Eqs.~(\ref{eq:alpha3M}) and (\ref{eq:alpha3S}) respectively. Notice that $\alpha_{\ell}^{(3,\bar{M})}$ does not depend on the EoS while $\alpha_{n,\ell}^{(3,\bar{S})}$ does. Notice also that these relations match those from~\cite{Stein:2014}, but also include corrections due to differential rotation to leading-order in $\gamma$.

For comparison, we temporarily restore units in the three-hair relations:
\be \label{3HRnonBar}
M_{\ell}+i\frac{q}{a}S_{\ell}= \bar{B}_{n,\lfloor\frac{\ell-1}{2}\rfloor}M_0\lb i q\rb^\ell\lb1+\gamma\,\alpha_{n,\ell}^{(3)}\rb +  \mathcal{O}(\gamma^2)\,,
\ee
where we have defined the piecewise function
\be
\alpha_{n,\ell}^{(3)}=\lcb
\begin{array}{ll}
\alpha_{\frac{\ell-2}{2}}^{(3,\bar{M})}&\,\text{even}\,\ell\nn\\
\alpha_{n,\frac{\ell-1}{2}}^{(3,\bar{S})}&\,\text{odd}\,\ell
\end{array}\right.\,,
\ee
and where $iq\equiv\sqrt{M_2/M_0}\,$, $a\equiv S_1/M_0$ and $\lfloor\cdot\rfloor$ is the floor function. The three-hair relations closely resemble the well-known no-hair relations for black holes: $M_\ell^{\BH}+iS_\ell^{\BH}=M_0(ia)^\ell$~\cite{Hansen:1974zz}.

Replacing the differential rotation parameter $\gamma$ with the next independent multipole moment $S_3$ turns the three-hair relations into a \textit{four-hair relation}. By taking the imaginary component of Eq.~\eqref{eq:3HR} with $\ell=1$, the expression can be solved for  $\gamma$ in terms of $\bar{M}_2$ and $\bar{S}_3$. The result simplifies into relations that only depend on the first four multipole moments:
\begin{align} \label{eq:4HR}
\bar{M}_{2\ell+2}+i\bar{S}_{2\ell+1} &= \bar{B}_{n,\ell}\bar{M}_2^{\ell}
\left[ \bar{M}_2 \left(1+\alpha_{n,\ell}^{(4,\bar{M})} \right)
\right. 
\nn \\
& \left. 
+ i \, \bar{S}_1 \left(1+\alpha_{n,\ell}^{(4,\bar{S})} \right) 
\right]+  \mathcal{O}(\gamma^2)\,,
\end{align}
where both $\alpha_{n,\ell}^{(4,\bar{M})}$ and $\alpha_{n,\ell}^{(4,\bar{S})}$ depend on $\bar{M}_2$ and $\bar{S}_3$ (but not on $\gamma$), and the moments were normalized by $M_0$ and $S_1$. These differential rotation corrections can be found in Eqs.~\eqref{eq:alpha4M} and~\eqref{eq:alpha4S} of Appendix~\ref{app:full}.

\subsection{Generic EoS Dependence}

\begin{figure}
\begin{center}
\includegraphics[width=\columnwidth,clip=true]{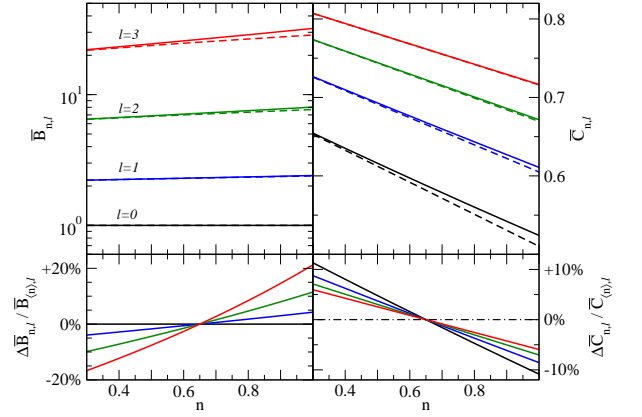}
\caption{\label{Coeff}(color online). $\bar{B}_{n,\ell}$ and $\bar{C}_{n,\ell}$ are plotted above for the first four $\ell$'s with $n\in[0.3,1]$. The solid lines represent the exact numerical results, while the dashed lines come from the perturbative approach. The bottom panels show the fractional difference of the semi-analytic results using a fiducial $n=0.65$.}
\end{center}
\end{figure}

We have found differential rotation-corrected three-hair relations and new four hair relations, but how EoS independent are these? Part of the EoS-dependence is absorbed in the three coefficients $\bar{A}_{n,\ell}$, $\bar{B}_{n,\ell}$, and $\bar{C}_{n,\ell}$, so we can analyze the variation of these with the polytropic index $n$. We will only look at  $\bar{B}_{n,\ell}$ and $\bar{C}_{n,\ell}$, since the EoS-dependence of $\bar{A}_{n,\ell}$ was already analyzed in~\cite{Stein:2014} in the uniformly rotating case, and found to lead to EoS variability of less than $10\%$. 

Figure~\ref{Coeff} shows the relative fractional differences of $\bar{B}_{n,\ell}$ and $\bar{C}_{n,\ell}$ computed with a polytropic EoS with $n \in [0.3,1]$ and a fiducial polytrope with $n = 0.65$. As found in~\cite{Stein:2014}, observe that the EoS variation in the uniform rotation coefficient, $\bar{B}_{n,\ell}$, increases as one increases $\ell$. However, $\bar{B}_{n,3}$ corresponds to the relation between $\bar S_7$ or $\bar M_8$ against $\bar M_2$. Therefore the $\sim 40\%$ variation in $\bar{B}_{n,3}$ between $n=0.3$ and $1$ are for less significant, higher order moments. Observe also that the EoS-dependence of the differential rotation correction $\bar{C}_{n,\ell}$ is smaller than that of the uniform rotation term $\bar{B}_{n,\ell}$ for large $\ell$, while the opposite is true for small $\ell$. However, recall from Eq.~\eqref{eq:4HR} that $\bar{C}_{n,\ell}$ is multiplied by $\gamma$, which is assumed to be much less than unity. Therefore, differential rotation does not destroy the EoS universality of the three-hair relation among lower multipole moments, at least when treated perturbatively and to Newtonian order in a weak-field expansion. 

We can gain some further mathematical understanding of this result by considering perturbations about an $n=0$ polytrope, for which we can solve the Lane-Emden (LE) equation exactly. Let us then set $n=\tilde{n}+\epsilon$, where $\tilde{n}$ is  a background polytropic index for which we assume we have a solution to the LE equation, and $\epsilon$ is a perturbation to $\tilde{n}$. The solution to the LE equation must then become $\vartheta=\tilde{\vartheta}+\epsilon \, \delta\vartheta+\mathcal{O}(\epsilon^2)$, where $\tilde{\vartheta}$ is the background solution to the LE equation and $\delta\vartheta$ is a perturbation. By substituting the perturbed variables into the LE equation and expanding to leading order in $\epsilon$, the perturbed LE equation becomes~\cite{1978SvA....22..711S,Bender:1989xe,Seidov:2004tt,Chatziioannou:2014}
\be\label{eq:pertLE}
2\tilde{\vartheta}\frac{d\delta\vartheta}{d\xi}+\tilde{\vartheta}\xi\frac{d^2\delta\vartheta}{d\xi^2}+\tilde{n}\tilde{\vartheta}^{\tilde{n}}\xi\delta\vartheta+\ln(\tilde{\vartheta})\tilde{\vartheta}^{\tilde{n}+1}\xi=0\,.
\ee
We require $\delta\vartheta(0)=0=\delta\vartheta^\prime(0)$, because $\tilde{\vartheta}$ already satisfies the initial conditions of the LE differential equation. 

The perturbed LE equation can be solved exactly when the background polytropic index is $\tilde{n} =0$. Doing so, we find~\cite{1978SvA....22..711S,Chatziioannou:2014}
\ba
\delta\vartheta^{(\tilde{n}=0)}&=&
\frac{\xi ^2}{18} \left[5-3 \ln \left(1-\frac{\xi ^2}{6}\right)\right]+3 \ln \left(1-\frac{\xi ^2}{6}\right)\nn\\
&+& \frac{4 \sqrt{6} \tanh ^{-1}\left(\xi/\sqrt{6}\right)}{\xi }-4\,.
\ea
From this solution, we can determine the perturbation to the location of the NS surface. Letting $\xi_1=\tilde{\xi}_1+\epsilon\, \delta\xi_1+\mathcal{O}(\epsilon^2)$, where $\tilde{\xi}_1=\sqrt{6}$ and requiring that $\vartheta(\xi_1)=0$, we find the perturbation to the NS surface to be~\cite{1978SvA....22..711S,Chatziioannou:2014}
\be
\delta\xi_1^{\tilde{n}=0}=\sqrt{6} \ln (4) - \frac{7}{\sqrt{6}}\,.
\ee

With all of these analytical results at hand, we can now evaluate the dependence of the EoS factors $\bar{B}_{n,\ell}$ and $\bar{C}_{n,\ell}$ as functions of $n$. First, we expand $\bar{B}_{0,\ell}=\tilde{\bar{B}}_{0,\ell}+\epsilon \, \delta \bar{B}_{0,\ell}+\mathcal{O}(\epsilon^2)$ and $\bar{C}_{0,\ell}=\tilde{\bar{C}}_{0,\ell}+\epsilon \, \delta \bar{C}_{0,\ell}+\mathcal{O}(\epsilon^2)$, noting that the background coefficients are given by
\begin{align}
\tilde{\bar{B}}_{0,\ell}&= 5^{\ell}\lsb\frac{15}{(2\ell+3)(2\ell+5)}\rsb\,,
\quad
\tilde{\bar{C}}_{0,\ell} = \frac{2\ell+5}{2\ell+7}\,.
\end{align}
The perturbation can then be solved for to find
\begin{align}
\frac{\delta \bar{B}_{0,\ell}}{\tilde{\bar{B}}_{0,\ell}}&=
\frac{-1}{15}\lsb15H\lb\ell+\frac{5}{2}\rb-6\ell-46+15\ln(4)\rsb\,,
\label{eq:Bpert}
\\
\frac{\delta \bar{C}_{0,\ell}}{\tilde{\bar{C}}_{0,\ell}}&=
\frac{-2}{2\ell+7}\,,
\label{eq:Cpert}
\end{align}
where $H(\ell)\equiv\sum_{k=1}^{\ell}1/k$ is the $\ell$th harmonic number. The same derivation for $\delta\bar{A}_{0,\ell}/\tilde{\bar{A}}_{0,\ell}$ was presented in~\cite{Chatziioannou:2014}, where the only change from Eq.~\eqref{eq:Bpert} is a factor of $-1/\ell$. The EoS-dependence of the coefficients can be analyzed by varying $\epsilon$. 

Figure~\ref{Coeff} compares these perturbative analytic solutions to the numerically calculated $\bar{B}_{n,\ell}$ and $\bar{C}_{n,\ell}$ in the top panels. The largest fractional difference between the numerical and perturbative calculations occur for $\bar{B}_{1,3}$ (a $10\%$ error) and for $\bar{C}_{1,0}$ (a $2.6 \%$ error). We then see that the perturbative calculation is more accurate for the differential rotation coefficient than for the uniform rotation one. By comparing Eqs.~\eqref{eq:Bpert} and~\eqref{eq:Cpert} we can also see how $\delta \bar{B}_{0,\ell}$ is typically larger than $\delta \bar{C}_{0,\ell}$ for large $\ell$, as observed in Figure~\ref{Coeff}.

\subsection{EoS Dependence in the Slow-Rotation Limit}

\begin{figure}
\begin{center}
\includegraphics[width=\columnwidth,clip=true]{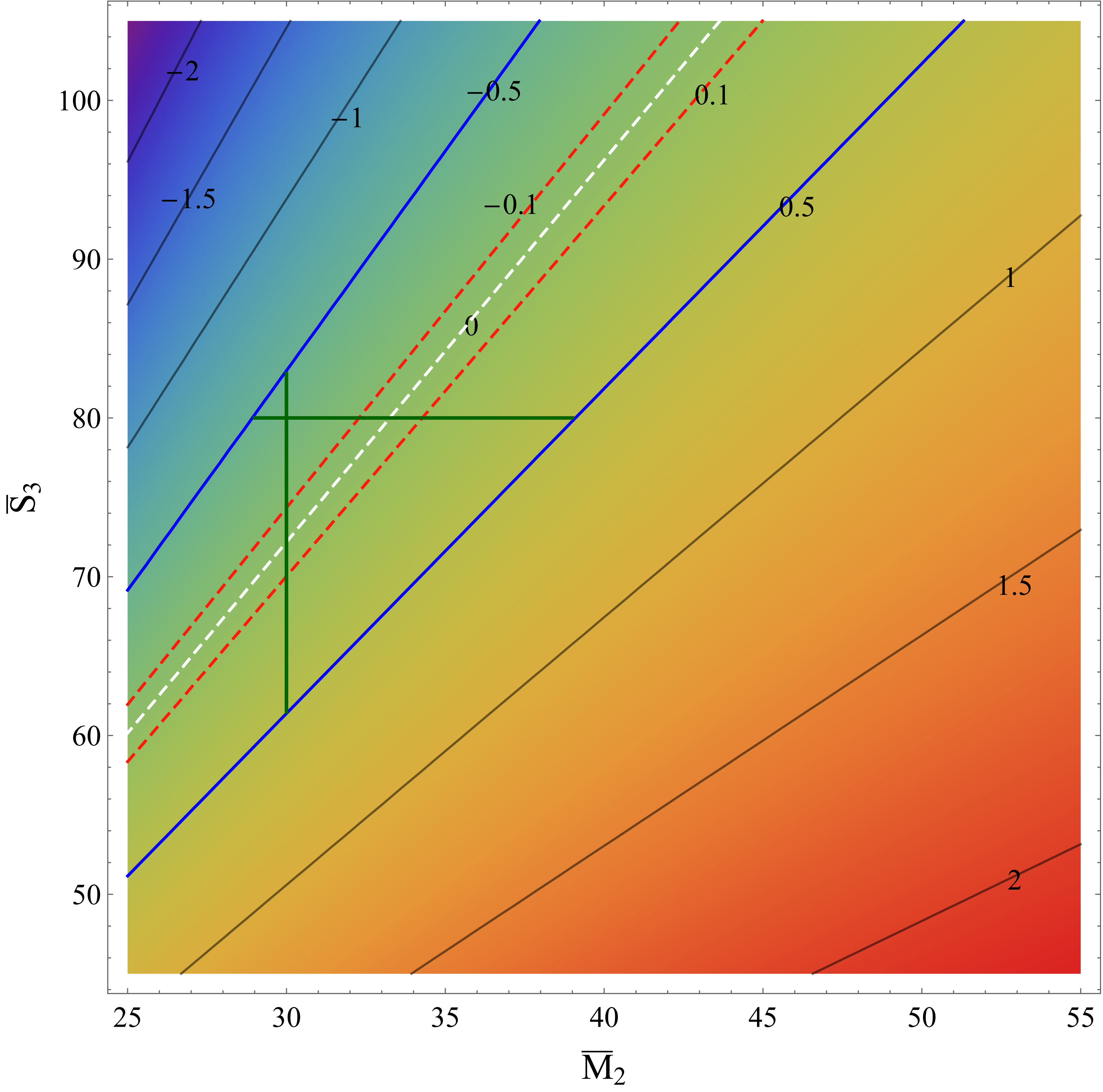}
\caption{\label{GammaPlot}(color online). Contour plot of $\gamma(\bar{M}_2,\bar{S}_3)$ under the slow-rotation limit. Observe how $\bar{M}_2$ and $\bar{S}_3$ specify the amount of differential rotation, and thus, if $|\gamma| < 0.1$ then only a subset of $\bar{M}_2$ and $\bar{S}_3$ are allowed. The contours are generated with a polytropic EoS with $n=1$ because values of $n<1$ yield less restrictive contours and thus a wider range of allowed $\bar{M}_2$ and $\bar{S}_3$. The white dashed zero-contour corresponds to the uniform rotation case, which shows a linear dependence between $\bar{M}_2$ and $\bar{S}_3$ that agrees with the three-hair relations for an $n=1$ polytrope. The green lines at $\bar M_2 = 30$ and $\bar S_5 = 80$ show the range of $\bar S_3$ and $\bar M_2$ used in Fig.~\ref{MomentsPlot} that is kept within $|\gamma| < 1/2$, demarcated by the blue lines.}
\end{center}
\end{figure}

\begin{figure*}
\begin{center}
\includegraphics[width=\columnwidth,clip=true]{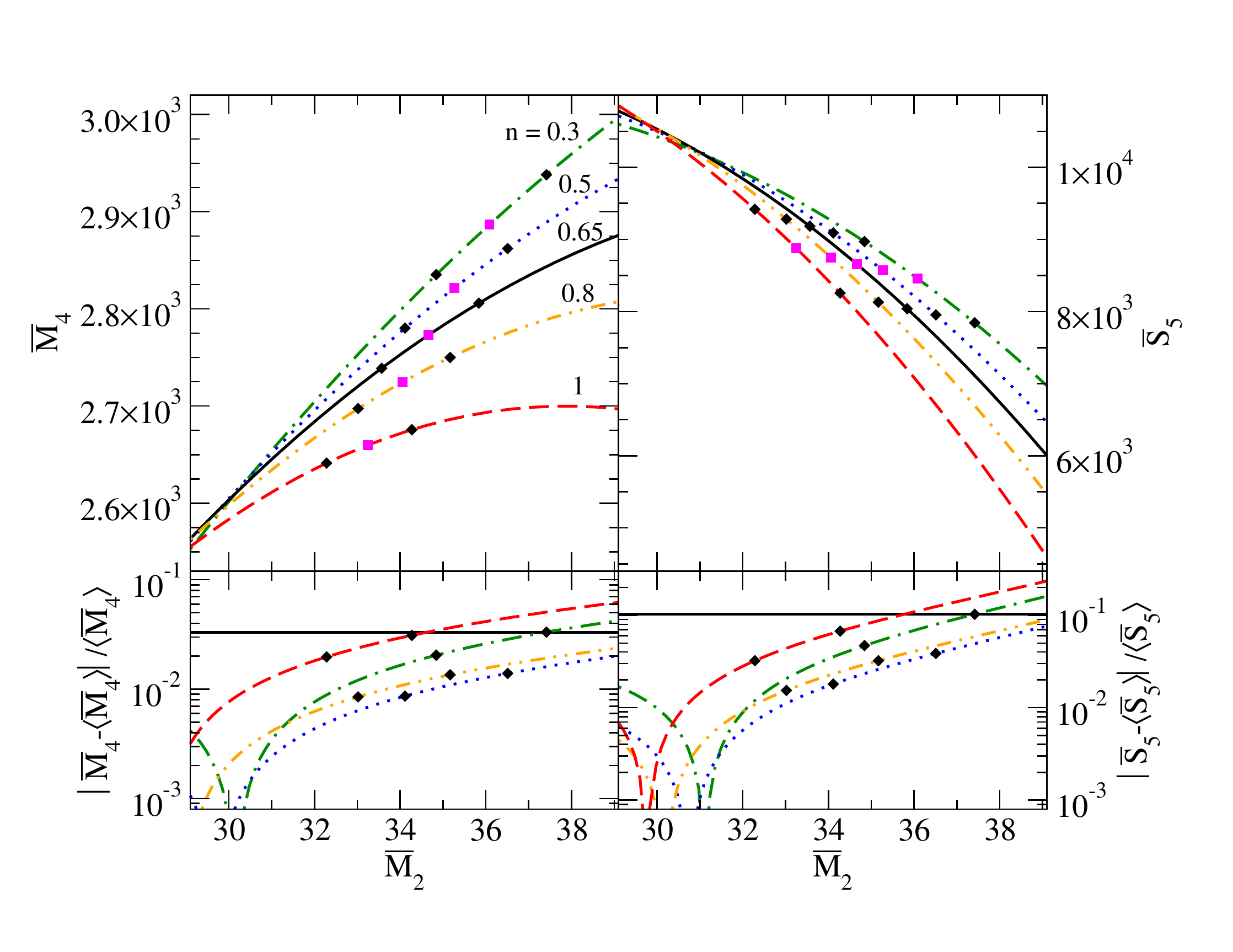}
\includegraphics[width=\columnwidth,clip=true]{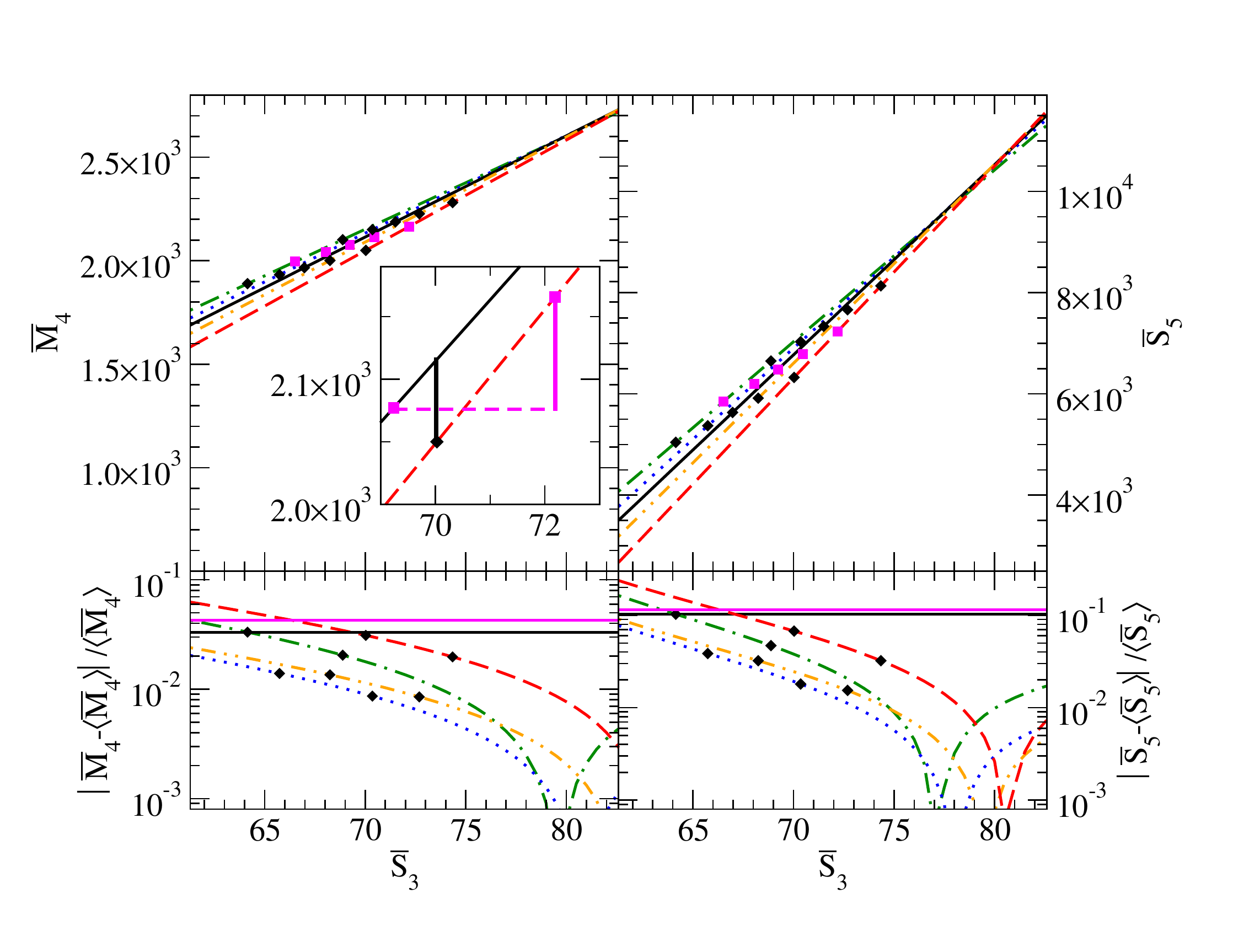}
\caption{\label{MomentsPlot}(color online). 
(Top panels) Four-hair relations for differentially rotating stars. The normalized moments $\bar{M}_4$ and $\bar{S}_5$ are plotted against $\bar{M}_2$ with $\bar S_3 = 80$ (left) and $\bar{S}_3$ with $\bar M_2 = 30$ (right) for various polytropic indices $n$ in the slow-rotation limit. Magenta squares represent the uniform rotation case ($\gamma=0$), while the black diamonds show where $\gamma=0.1$ for a given value of $n$. The fiducial polytropic index of $n=0.65$ is plotted as a solid black line. The inset zooms into a region around $\bar{S}_3 = 71$ and $\bar M_4 = 2.1\times10^3$, where the vertical magenta (black) line corresponds to the difference in $\bar M_4$ between an $n=0.65$ and $n=1$ polytrope for uniformly (differentially) rotating stars with fixed $\bar M_{2}$. Observe that the latter is smaller than the former when $|\gamma| < 0.1$ with an $n=1$ polytrope. (Bottom panels) The fractional difference of each of the four-hair relations relative to that of a fiducial polytrope ($n=0.65$). Observe that the maximum EoS variation is roughly 7\% for $\bar{M}_4$ and 20\% for $\bar{S}_5$, but this goes down to 3\% for $\bar{M}_4$ and 10\% for $\bar{S}_5$ in the $|\gamma| < 0.1$ region marked by the horizontal black lines. The horizontal magenta lines show the maximum fractional difference in the three-hair relations for uniformly rotating stars.}
\end{center}
\end{figure*}

In the previous subsection, we investigated the EoS-dependence of the differential rotation correction $\bar{C}_{n,\ell}$ of Eq.~\eqref{eq:C-def}, which clearly does not depend on the angular integrals $\tilde{I}_{\ell,k}$ of Eq.~\eqref{eq:Intt}. These angular integrals, however, enter the three- and four-hair relations in Eqs.~\eqref{eq:3HR} and~\eqref{eq:4HR} through the $\alpha_{n,\ell}^{(3,4,\bar{M},\bar{S})}$ coefficients. We see that they depend on the Legendre coefficients $\alpha_{2 j}$ that characterize the deformations of the stellar surface. Appendix~\ref{app:shape} derives such a deformation and shows how to solve for the Legendre coefficients in the slow-rotation limit, i.e.~expanding in $\chi := S_1/M_0^2 \ll 1$. The first three $\alpha_{2j}$'s are given explicitly in Eq.~\eqref{eq:alphas}. 

In the slow-rotation approximation, we can then compute the first few multipole moments from Eqs.~\eqref{eq:M-fin} and~\eqref{eq:S-fin}. The first mass moments are
\begin{align}
\label{eq:M2-slow}
M_2 &=-\frac{a_1^2 e^2 M_0  \mathcal{R}_{n,2}}{3 \xi_1^4 |\vartheta^\prime(\xi_1)|} \lb1-\gamma\frac{16}{5}\rb +  \mathcal{O}(\gamma^2)\,,
\\
M_4&=\frac{a_1^4 e^4 M_0 \mathcal{R}_{n,4}}{5 \xi _1^6 |\vartheta^\prime(\xi_1)|} \lb1- \frac{\gamma}{e^2} \frac{8}{15}\rb +  \mathcal{O}(\gamma^2)\,,
\end{align}
and the first current moments are
\begin{align}
S_1 &=\frac{2 a_1^2 M_0 \Omega_c \mathcal{R}_{n,2}}{3 \xi_1^4 |\vartheta^\prime(\xi_1)|} \lb1-\gamma \frac{8}{15\xi_1^2}\frac{\mathcal{R}_{n,4}}{\mathcal{R}_{n,2}}\rb +  \mathcal{O}(\gamma^2)\,,
\\
S_3 &=-\frac{2 a_1^4 e^2 M_0 \Omega _c \mathcal{R}_{n,4}}{5 \xi_1^6 |\vartheta^\prime(\xi_1)|}\left(1
- \frac{\gamma}{e^2} \frac{24 }{49 \xi_1^2 }\frac{\mathcal{R}_{n,6}}{\mathcal{R}_{n,4}}\right) +  \mathcal{O}(\gamma^2)\,,
\\
\label{eq:S5-slow}
S_5&=\frac{2 a_1^6 e^4 M_0 \Omega _c \mathcal{R}_{n,6}}{7 \xi_1^8 |\vartheta^\prime(\xi_1)|}\left[1
- \frac{\gamma}{e^2} \frac{8}{11} \left(\frac{3}{5}
+\frac{5}{3 \xi _1^2 }\frac{\mathcal{R}_{n,8}}{\mathcal{R}_{n,6}}\right)\right] \nn \\
&+  \mathcal{O}(\gamma^2)\,.
\end{align}
Notice that the differential rotation corrections to $M_2$ and $S_1$ are proportional to $\gamma$ while those to $M_4$, $S_3$ and $S_5$ are proportional to $\gamma/e^2$. This does not mean that our expressions are divergent in the $e^{2} \to 0$ limit; if one expands out the expressions above, one finds that they are all finite in this limit. The $\gamma/e^{2}$ dependence, however, does mean that our double expansion of slow-rotation and small differential rotation is only valid when $|\gamma |\ll e^2 \ll 1$. Notice that $S_3 \propto \Omega^{3} + \gamma \, \Omega$, and thus, it is non-vanishing to linear order in spin in differentially rotating stars, as already implied in Refs.~\cite{Passamonti:2005cz,Stavridis:2007xz,Passamonti:2007td,Chirenti:2013xm} for a j-constant law in full GR. Notice also that $M_{4} \propto \Omega^{4} + \gamma \, \Omega^{2}$ and $S_{5} \propto \Omega^{5} + \gamma \, \Omega^{3}$, and thus these moments are non-zero at second and third order in spin respectively for differentially rotating stars. 

From Eqs.~\eqref{eq:M2-slow}--\eqref{eq:S5-slow}, we can further compute the four-hair relation in the slow-rotation limit to find
\be
\bar{M}_4=\bar{M}_2^2 \bar{B}_{n,1}\lsb1+\frac{49}{45}\frac{ \left(\bar{S}_3-\bar{M}_2 \bar{B}_{n,1}\right)}{ \bar{M}_2 \bar{B}_{n,1} \bar{C}_{n,1}}\rsb+  \mathcal{O}(\gamma^2)\,,
\ee
\ba
\bar{S}_5 &=& \bar{M}_2^2 \bar{B}_{n,2}\lsb1+\frac{49}{45}\frac{\left(9+25 \bar{C}_{n,2}\right)}{11}\frac{  \left(\bar{S}_3-\bar{M}_2 \bar{B}_{n,1}\right)}{\bar{M}_2 \bar{B}_{n,1} \bar{C}_{n,1}}\rsb \nn \\
&+&  \mathcal{O}(\gamma^2)\,.
\ea
Observe how these relations depend on $\bar{C}_{n,\ell}$ and $\bar{B}_{n,\ell}$, as discussed in the previous subsection.

Before we can plot these four-hair relations in the slow-rotation limit, we must choose a range of values for $\bar{M}_{2}$ and $\bar{S}_{3}$. In doing so, however, one must be careful to pick a range that satisfies $|\gamma| \ll 1$ so that the perturbative differential rotation approximation is not violated. The dependence of $\bar{M}_2$ and $\bar{S}_3$ on $\gamma$ is shown through contours in Fig.~\ref{GammaPlot}. Observe that only a subset of values of $(\bar{M}_{2},\bar{S}_{3})$, shown with dotted red lines, lead to $|\gamma| < 0.1$. This range of values is in fact the smallest range possible, since the contours were constructed with a polytropic EoS with $n=1$ and other values of $n<1$ would allow for a larger range. 

\begin{figure*}
\begin{center}
\includegraphics[width=270pt,clip=true]{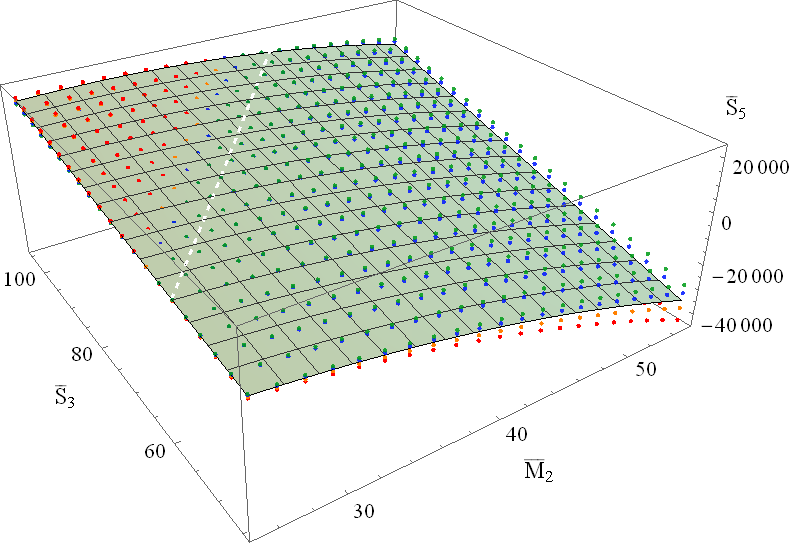}
\includegraphics[width=225pt,clip=true]{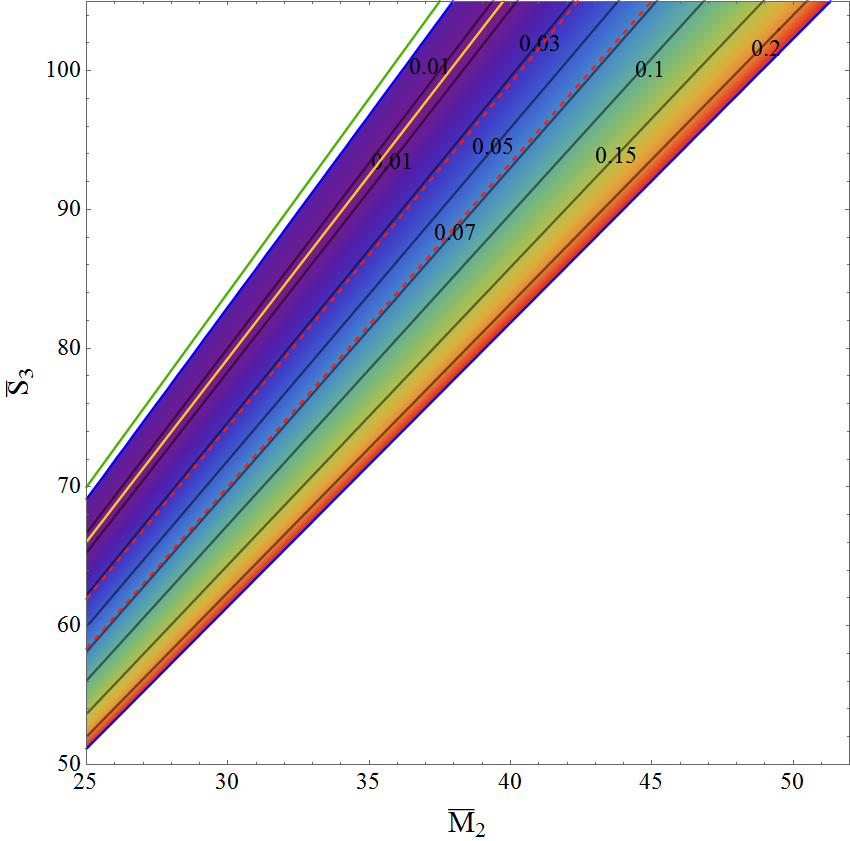}
\caption{\label{S5Plots}(color online). 
Same as Fig.~\ref{M4Plots} but for $\bar{S}_5$ as a function of $\bar M_2$ and $\bar S_3$. The yellow (green) line in the right panel shows where the fractional difference of $\bar{S}_{5}$ computed with an $n= 0.3$ ($n=1$) polytrope vanishes with respect to the fiducial $n=0.65$ one. Observe that the maximum fractional difference is always less than $\sim$ $20\%$ for all values of $\bar{S}_3$ and $\bar{M}_2$ that satisfy $|\gamma| < 1/2$, while it is only between $3\%$ and $7\%$ in the $|\gamma| < 0.1$ region (dotted red lines).}
\end{center}
\end{figure*}

With the range of allowed values explored, we now plot the four-hair relations in the slow-rotation limit in two dimensions (Fig.~\ref{MomentsPlot}) and in three-dimensions (Figs.~\ref{M4Plots} and~\ref{S5Plots}). The left panel of Fig.~\ref{MomentsPlot} shows the $\bar{M}_{4}$-$\bar{M}_{2}$ relation and the $\bar{S}_{5}$-$\bar{M}_{2}$ relation for a fixed value of $\bar{S}_{3} = 80$ over a range of values of $\bar{M}_{2}$ that guarantee that $|\gamma| < 1/2$ (the horizontal green line in Fig.~\ref{GammaPlot}). Similarly, the right panel of Fig.~\ref{MomentsPlot} shows the $\bar{M}_{4}$-$\bar{S}_{3}$ relation and the $\bar{S}_{5}$-$\bar{S}_{3}$ relation for a fixed value of $\bar{M}_{2} = 30$ over a range of values of $\bar{S}_{3}$ that again satisfy $|\gamma| < 1/2$  (the vertical green line in Fig.~\ref{GammaPlot}). Observe how the relative fractional difference (bottom panels) shows a weak EoS-dependence, i.e.~a weak dependence on $n$, relative to a fiducial polytropic index of $n = 0.65$. The maximum fractional differences in the $\bar{M}_{4}$-$\bar{M}_{2}$ and $\bar{M}_{4}$-$\bar{S}_{3}$ relations is $\sim 7\%$, while in the $\bar{S}_{5}$-$\bar{M}_{2}$ and $\bar{S}_{5}$-$\bar{S}_{3}$ relations it is $\sim 20 \%$. But notice that these differences decrease to $\sim 3\%$ and $\sim 10\%$ respectively for $|\gamma| < 0.1$, as shown by the horizontal black lines; again, differential rotation does not spoil the approximate EoS universality.  

Let us now compare the EoS variation in the four-hair relations for differentially rotating stars to that in the three-hair relations for uniformly rotating ones. The latter is simple to obtain: we simply take the difference between $\bar{M}_{4}$ using $n=0.65$ and $n=1$ polytropes with a fixed value of $\bar M_2$, shown by the vertical magenta line between the magenta squares in the inset of the top, right panel of Fig.~\ref{MomentsPlot}, a zoom to the region around $\bar{S}_3 = 71$ and $\bar M_4 = 2.1\times10^3$. The EoS variation in the differentially rotating case, however, is more difficult to obtain because it depends both on $\bar{M}_{2}$ and $\bar{S}_{3}$. For example, if we fix $\bar{M}_{2}$, as we do in the top, right panel of Fig.~\ref{MomentsPlot}, the EoS variation between $n=0.65$ and $n=1$ polytropes is represented by the distance between the black solid line and the red dashed line, which clearly depends on the value of $\bar{S}_{3}$ chosen. Let us fix the value of $\bar{S}_{3}$ such that the EoS variation is largest within the range of values of $\bar{S}_{3}$ that corresponds to $|\gamma| < 0.1$ for an $n=1$ polytrope, i.e.~the largest vertical distance between the black solid line and the red dashed line within the region delimited by the black dots of the red dashed line. 

With this at hand, we can now compare the EoS variations, shown in the bottom, right panels of Fig.~\ref{MomentsPlot}. The horizontal magenta line shows the relative fractional difference in the uniform rotation case, while the black solid line shows the maximum, relative fractional difference in the differential rotation case for $|\gamma| < 0.1$. Observe that the universality improves slightly when one includes differential rotation, i.e.~the horizontal magenta line is slightly above the horizontal black line. This result is robust to the choices made above, since if we had picked different values of $\bar{S}_{3}$ (within $|\gamma|<0.1$ for an $n=1$ polytrope) to compute the EoS variation in the differential rotation case, the latter would have been even smaller, i.e.~the horizontal black line would have been farther down. The reason for this is that differential rotation introduces a new degree of freedom, encoded in $\bar{S}_{3}$ in the four-hair relations, which allows the EoS variation to explore a new direction (the vertical direction in the inset of Fig.~\ref{MomentsPlot}). For small values of $|\gamma|$, there are values of $\bar{S}_{3}$, different from those required in uniform rotation for a fixed $\bar{M}_{2}$, that lead to a smaller degree of variation. Of course, if one considers $|\gamma| > 0.1$, the EoS variation in the four-hair relations eventually exceeds that in the three-hair relations for uniformly rotating stars. Similar features can be seen for the $\bar S_5$-$\bar S_3$ case with fixed $\bar M_2$. For the $\bar S_5$-$\bar M_2$ case with fixed $\bar S_3$, differential rotation deteriorates the universality from the uniform rotation case slightly, but the latter does not correspond to the universality of the original three-hair relations in~\cite{Stein:2014}, since that fixes $\bar M_2$ instead of $\bar S_3$.

Three-dimensional plots of the four-hair relations in the slow-rotation limit are shown in Figs.~\ref{M4Plots} and~\ref{S5Plots}. The left panel of Fig.~\ref{S5Plots} shows how the $\bar{S}_{5}$-$\bar{M}_{2}$-$\bar{S}_{3}$ relation exists on an \emph{invariant} plane that is approximately insensitive to the EoS. The right panel shows this insensitivity in more detail, by plotting the maximum fractional difference in $\bar{S}_{5}$ in the part of the $\bar{S}_{3}$-$\bar{M}_{2}$ plane that satisfies $|\gamma | < 1/2$. The yellow (green) solid line shows where the fractional difference of $\bar{S}_{5}$ computed with an $n=0.3$ ($n=1$) polytrope and a fiducial $n=0.65$ polytrope vanishes. Observe how these lines lie inside the purple region, where the EoS variation is smallest. Observe again how the maximum EoS variability is always less than $\sim 20\%$ inside the $|\gamma| < 1/2$ region, but it is less than $\sim 7\%$ in the $|\gamma| < 0.1$ region. 

\section{Discussion}
\label{sec:disc}

We have studied the effect of differential rotation in the approximately universal three-hair relations for neutron stars. To allow for a semi-analytic treatment, we used a leading-order weak-field, perturbative differential rotation and slow-rotation expansion. Under these approximations, our results show that differential rotation affects the three-hair relations only mildly, without spoiling the universality found in uniform rotation for sufficiently weak differential rotation. We also found that differential rotation introduces a new parameter into the three-hair relations that quantifies the degree of differential rotation. We eliminated this parameter in favor of the next non-vanishing multipole moment (the mass-current octopole) to derive new four-hair relations, which typically have smaller EoS variation for small differential rotation than the three-hair relations for uniformly rotating stars. 

In order to proceed with our analysis, we chose a generic class of rotation laws that encompasses the j- and v- constant laws, as well as Keplerian and HMNS rotation laws, all of which are identical to linear order in small differential rotation. Although other classes of rotation laws could have been used to model more realistic differential rotation curves, we expect such laws to only induce corrections to the three-hair relations of $\mathcal{O}(\gamma)$ relative to the uniformly rotating case, as long as one imposes the small differential rotation approximation. Therefore, provided that the amount of differential rotation is small, the results presented here should also be valid qualitatively for other classes of rotation laws.

Our results suggest that the approximate EoS universality may survive the presence of differential rotation, but a much more detailed numerical analysis would be required to confirm this. In particular, one could consider the multipole moments in full GR (without a weak-field expansion), for rapidly rotating stars (without assuming slow-rotation) and for large differential rotation, such as that produced in the remnant shortly after a supernova or a binary merger. Doing so, one could extend the work of~\cite{Martinon:2014uua} on PNSs to consider differential rotation and universality. One could also study whether the approximate universality found in the new four-hair relations could possibly be improved through a different choice of normalization, which was recently found for the three-hair relations in the absence of differential rotation~\cite{Majumder:2015kfa}. 

\acknowledgements
We would like to thank Leonardo Gualtieri for useful comments and suggestions. We also acknowledge support from NSF CAREER Award PHY-1250636.

\appendix
\section{Stellar Shape Deformation for Differentially Rotating Stars}
\label{app:shape}

We here extend the analysis of Sec.~2.4.4 in~\cite{2014grav.book.....P} to derive the stellar shape of differentially rotating stars to leading (Newtonian) order in a weak-field expansion. We work in a perturbative differential rotation and a slow-rotation approximation, and thus, we expand the shape about a non-rotating background. The background satisfies a mass conservation law and a hydrostatic equilibrium equation:
\be
\label{eq:hydrostatic-bg}
m(r) = 4 \pi \int_0^R \rho(r) r^2 dr\,, \quad \frac{dp}{dr} = - \rho(r) \frac{m(r)}{r^2}\,,
\ee
where $R$ is the stellar radius in the non-rotating configuration.

Let us next look at the perturbed equations. Focusing on a generic, non-axisymmetric perturbation, the perturbed mass conservation law and hydrostatic equilibrium equation are given by 
\be
\label{eq:hydrostatic-pert}
\delta \rho = - \xi^j \partial_j \rho\,, \quad \frac{\delta \rho}{\rho^2} \partial_j p - \frac{1}{\rho} \partial_j \delta p + \partial_j (\delta U + V)=0\,,
\ee
where $\delta p$, $\delta \rho$ and $\delta U$ are the perturbations to the pressure, the density and the gravitational potential respectively. We have also introduced $\xi^i$ as the displacement of a fluid element in the perturbed configuration relative to the background and $V$ as the potential that drives the perturbation (namely, the centrifugal potential).

Let us now decompose the perturbed quantities in spherical harmonics:
\ba
\xi^r &=& \sum_{\ell m} r f_{\ell m}(r) Y_{\ell m} (\theta, \phi)\,, \\
\delta p &=& \sum_{\ell m} p_{\ell m}(r) Y_{\ell m} (\theta, \phi)\,, \\
\delta U &=& \sum_{\ell m} U_{\ell m}(r) Y_{\ell m} (\theta, \phi)\,, \\
\label{eq:V-deomp}
V &=& \sum_{\ell m} V_{\ell m}(r) Y_{\ell m} (\theta, \phi)\,.
\ea
Substituting the above decomposition into the perturbed mass conservation law in Eq.~\eqref{eq:hydrostatic-pert} and using the EoS and the background hydrostatic equilibrium equation in Eq.~\eqref{eq:hydrostatic-bg}, one finds 
\be
\label{eq:mass-con-pert-deomp}
p_{\ell m} = \frac{\rho m}{r} f_{\ell m}\,.
\ee
Similarly, from the angular component of Eq.~\eqref{eq:hydrostatic-pert}, one finds 
\be
\label{eq:hydrostatic-pert-deomp}
p_{\ell m} = \rho (U_{\ell m} + V_{\ell m})\,.
\ee
Combining Eqs.~\eqref{eq:mass-con-pert-deomp} and~\eqref{eq:hydrostatic-pert-deomp}, one finds
\be
\label{eq:fUV}
\frac{m}{r} f_{\ell m} = U_{\ell m} + V_{\ell m}\,.
\ee

The goal of this appendix is to derive the stellar shape deformation,
\be
\label{eq:deltaR}
\delta R = \sum_{\ell m} R \, f_{\ell m} (R) Y_{\ell m} (\theta, \phi)\,,
\ee
and thus, we need to derive a relation between $f_{\ell m}$ and $V_{\ell m}$. From the perturbed Poisson equation in the exterior region, one finds
\be
U_{\ell m}(r) = \frac{4\pi}{2 \ell +1} \frac{M_{\ell m}}{r^{\ell +1}}\,,
\ee
where $M_{\ell m}$ correspond to the stellar mass multipole moments. For an axisymmetric configuration, $M_{\ell 0}$ agrees with $M_\ell$ in Eq.~\eqref{eq:Ml-integral} modulo a constant factor.
Equation~\eqref{eq:fUV} at $r=R$ then becomes
\be
\frac{M}{R} f_{\ell m}(R) = \frac{4\pi}{2\ell +1} \frac{M_{\ell m}}{R^{\ell+1}} + V_{\ell m} (R)\,,
\ee
and a similar equation holds for the first derivative of Eq.~\eqref{eq:Ml-integral}. 
Eliminating $M_{\ell m}$ from these two equations, one finds
\be
\label{eq:fV}
f_{\ell m} (R) = \frac{h_\ell}{2 \ell +1} \frac{R}{M} \left[ (\ell + 1) V_{\ell m}(R) + R V'_{\ell m}(R) \right]\,,
\ee
where we introduced the \emph{surficial Love number} $h_\ell$, defined by
\be
h_\ell \equiv \frac{2 \ell+1}{\ell + \eta_\ell(R)}\,, 
\ee
which depends on the EoS, and the Radau function
\be
\eta_\ell (r) \equiv \frac{r f'_{\ell m}}{f_{\ell m}}\,.
\ee

Let us now focus on an axisymmetric rotational perturbation and consider the $m=0$ mode only. Expanding in $\chi \equiv S_{1}/M_{0}^{2} \ll 1$ and in $|\gamma| \ll 1$, the centrifugal potential is given by
\ba
\label{eq:centrigufal-pot}
V(r,\theta) &=&  \frac{1}{2} \Omega(r,\theta)^2 r^2 \sin^2 \theta + \mathcal{O}\left( \chi^4 \right)\,, \nn \\
&=& \frac{1}{3} \Omega_c^2 r^2 \left[ 1 + P_2(\mu) \right] \nn \\
& & - \frac{8}{15} \gamma \frac{\Omega_c^2 r^4}{a_1^2} \left[ 1 -\frac{10}{7} P_2(\mu) +\frac{3}{7} P_4(\mu) \right] \nn \\
& &+ \mathcal{O} \left( \chi^4, \gamma^2 \right)\,, 
\ea
where recall that $\mu \equiv \cos \theta$ and we used Eq.~\eqref{eq:Omega} for $\Omega(r,\theta)$. Comparing Eqs.~\eqref{eq:V-deomp} and~\eqref{eq:centrigufal-pot}, one can read off $V_{\ell 0}$ for the rotational perturbation:
\begin{align}
V_{00} &= \frac{2\sqrt {\pi }}{3} {\Omega_c}^{2}{r}^{2} \left( 1 - \frac{8}{5} \gamma {\frac {{r}^{2}}{{a_1}^{2}}} \right)+ \mathcal{O} \left( \chi^4,  \gamma^2 \right)\,, \\
V_{20} &= - \frac{2 \sqrt{5 \pi}}{15}{\Omega_c}^{2}{r}^{2} \left( 1 -  \frac{16}{7} \gamma {\frac {{r}^{2}}{{a_1}^{2}}} \right)+ \mathcal{O} \left( \chi^4,  \gamma^2 \right)\,, \nn \\
\\
V_{40} &= - \frac{16 \sqrt{\pi}}{105} \gamma {\frac {{\Omega_c}^{2}{r}^{4}}{{a_1}^{2}}} + \mathcal{O} \left( \chi^4,  \gamma^2 \right)\,.
\end{align}
From these expressions, one can obtain the perturbations $f_{\ell m}(R)$, and from that the stellar shape deformation using Eqs.~\eqref{eq:deltaR} and~\eqref{eq:fV}. Re-expressing $R$ and $\Omega_c^2$ in terms of $a_1$ and $e^2$, one finds
\ba
\delta R &=& a_1 \left\{ 1 - \frac{1}{6} \left[ 1 + 2 P_2(\mu) \right] e^2 \right. \nn \\
& & \left. + \frac{16}{15} \gamma \left[ P_2 (\mu) - \frac{3\, h_4}{14\, h_2} P_4(\mu) \right] e^2\right\}   + \mathcal{O} \left( e^4,  \gamma^2 \right)\,. \nn \\
\label{eq:R-def}
\ea
The uniform rotation part in the above equation corresponds to the spheroidal shape deformation, while the $\mathcal{O}(\gamma)$ part is a correction due to differential rotation. 

Once the stellar shape deformation has been found, one can read off the Legendre coefficients $\alpha_{2 j}$. Comparing Eq.~\eqref{eq:R-def} to Eq.~\eqref{eq:Intt}, one finds
\ba
\label{eq:alpha}
\alpha_0 &=& \mathcal{O} \left( e^4 \right)\,, \quad \alpha_2 = \frac{16}{15} e^2 + \mathcal{O} \left( e^4 \right)\,, \nn \\
\alpha_4 &=& - \frac{8 \, h_4}{35 \, h_2} e^2 + \mathcal{O} \left( e^4 \right)\,.
\ea
Since $h_\ell$ enters only at $\mathcal{O}(\gamma)$, we need to evaluate it, and thus the Radau function, for a uniformly rotating star. The Radau function satisfies the Radau equation, which for uniform rotation is
\be
r \eta_\ell' + \eta_\ell (\eta_\ell -1) +8 \pi \frac{\rho r^3}{m} (\eta_\ell +1) - \ell (\ell +1) = 0\,,
\ee
with the boundary condition $\eta_\ell (0) = \ell -2$. For an $n=0$ polytropic EoS, and taking into account the density discontinuity at the surface, one finds 
\be
h_\ell ^{(n=0)} = \frac{2 \ell+1}{2(\ell -1)}\,.
\ee
Substituting this into Eq.~\eqref{eq:alpha}, one obtains $\alpha_{\ell}$ for an $n=0$ polytrope:
\begin{align}
\label{eq:alphas}
\alpha_{0} &= {\cal{O}}(e^{4})\,, \qquad\alpha_{2}=\frac{16}{15}e^2 + {\cal{O}}(e^{4})\,, \nn
\\
\alpha_{4} &=-\frac{24}{175}e^2 + {\cal{O}}(e^{4})\,.
\end{align}

\section{Detailed Expressions for the Multipole Moments for Differentially Rotating Stars}
\label{app:full}

\bw
We here present detailed and generic expressions for the mass and mass-current multipole moments with differential rotation within the perturbative differential rotation and weak-field approximations. Normalizing Eqs.~\eqref{eq:M-fin} and~\eqref{eq:S-fin}, we find
\ba \label{eq:M2lp2}
\bar{M}_{2\ell+2}&=&
\frac{(-1)^{\ell+1}}{4^{\ell+1}}
\frac{3^{2\ell+2}}{\Omega _c^{2\ell+2}}
\frac{\xi_1^{6 \ell+4}}{a_1^{2\ell+2}}
\frac{I_{0,3}^{2\ell+1}}{\delta I_{1,3}^{2\ell+2}}
\frac{|\vartheta^\prime(\xi_1)|^{2\ell+1}}{I_{2\ell+2,3}^{-1}}
\frac{\mathcal{R}_{n,2\ell+2}}{\mathcal{R}_{n,2}^{2\ell+2}}
\left[1+\gamma\left(
3(2\ell+1)\frac{\tilde{I}_{0,2}}{I_{0,3}}
-10(\ell+1)\frac{\delta\tilde{I}_{1,2}}{\delta I_{1,3}}
+(2\ell+5)\frac{\tilde{I}_{2\ell+2,2}}{I_{2\ell+2,3}} \right.\right.\nn\\
&+&\left.\left.\frac{4}{3}(\ell+1)\frac{\delta I_{1,5} -\delta\bar{I}_{1,5}}{\delta I_{1,3}} \frac{\mathcal{R}_{n,4}}{\xi_1^2 \mathcal{R}_{n,2}}
\right)\right]+  \mathcal{O}(\gamma^2)\,,
\ea
\ba \label{eq:S2lp1}
\bar{S}_{2\ell+1}&=&
\frac{(-1)^\ell}{4^\ell}
\frac{3^{2\ell+1}}{\Omega _c^{2\ell}}
\frac{2\ell+1}{4\ell+3}
\frac{\xi_1^{6\ell}}{a_1^{2\ell}}
\frac{I_{0,3}^{2\ell}}{\delta I_{1,3}^{2\ell+1}}
\frac{|\vartheta^\prime(\xi_1)|^{2\ell}}{\delta I_{2\ell+1,3}^{-1}}
\frac{\mathcal{R}_{n,2\ell+2}}{\mathcal{R}_{n,2}^{2\ell+1}}
\left\{1+\gamma\left[
6\ell\frac{\tilde{I}_{0,2}}{I_{0,3}}
-5(2\ell+1)\frac{\delta\tilde{I}_{1,2}}{\delta I_{1,3}}
+(2\ell+5)\frac{\delta\tilde{I}_{2\ell+1,2}}{\delta I_{2\ell+1,3}}
\right.\right.\nn\\
&+&\left.\left.
2(2\ell+1)\left(
\frac{\delta I_{1,5}-\delta\bar{I}_{1,5}}{3\delta I_{1,3}}
\frac{\mathcal{R}_{n,4}}{\xi_1^2 \mathcal{R}_{n,2}}
-\frac{\ell+1}{4\ell+3}
\frac{\delta I_{2\ell+1,5}-\delta\bar{I}_{2\ell+1,5}}{\delta I_{2\ell+1,3}}
\frac{\mathcal{R}_{n,2\ell+4}}{\xi_1^2 \mathcal{R}_{n,2\ell+2}}
\right)\right]\right\}+  \mathcal{O}(\gamma^2)\,.
\ea

These multipole moments are the basis for the universal relations that lead to the three- and four-hair relations in the presence of differential rotation. The first and second universal relations of Eqs.~\eqref{eq:FirstUniv} and~\eqref{eq:SecondUniv} are corrected due to differential rotation through the coefficients $\alpha^{(1,2)}_{n,\ell}$, which are explicitly
\ba \label{eq:alpha1}
\alpha_{n,\ell}^{(1)}&=& \frac{1}{2\sqrt{1-e^2}} \left[5 \delta\tilde{I}_{1,2}+15\frac{\tilde{I}_{2,2}}{e^2} +(-1)^{\ell+1}(2 \ell+3) (2 \ell+5) \left(\frac{\tilde{I}_{2 \ell+2,2}}{e^{2\ell+2}} +\frac{2\ell+1}{4\ell+3}\frac{\delta\tilde{I}_{2 \ell+1,2}}{e^{2\ell}}\right)\right]
 -\frac{8}{15}\bar{C}_{n,0}\nn\\
&+&\frac{4(-1)^{\ell+1} (\ell+1) (2 \ell+1) \left(2 e^2+\ell\right)}{(2\ell+5) (4 \ell+3)e^2} \bar{C}_{n,\ell}\,,
\ea
\ba \label{eq:alpha2}
\alpha_{n,\ell}^{(2)}&=& \frac{1}{2\sqrt{1-e^2}}
\left[5\frac{\ell+1}{\ell} \delta\tilde{I}_{1,2}
-3\tilde{I}_{0,2}
+(-1)^{\ell+1}(2\ell+3)(2\ell+5)\left( \frac{\tilde{I}_{2\ell+2,2}}{e^{2\ell+2}}
+\frac{\ell+1}{\ell}\frac{2\ell+1}{4\ell+3} \frac{\delta \tilde I_{2 \ell+1,2}}{e^{2\ell}}\right)\right]\nn\\
&-&\-\frac{8}{15}\frac{\ell+1}{\ell} \bar{C}_{n,0}
+\frac{4(-1)^{\ell+1}(\ell+1)(2\ell+1) \left(2e^2+\ell\right)}{(2\ell+5)(4\ell+3)e^2}\frac{\ell+1}{\ell} \bar{C}_{n,\ell}\,.
\ea
The three- and four-hair relations of Eqs.~\eqref{eq:3HR} and~\eqref{eq:4HR} in the presence of differential rotation are corrected by the coefficients $\alpha_{\ell}^{(3,\bar{M},\bar{S})}$ and $\alpha_{\ell}^{(4,\bar{M},\bar{S})}$, which are explicitly
\ba \label{eq:alpha3M}
\alpha_{\ell}^{(3,\bar{M})}=
\frac{1}{2 \sqrt{1-e^2}}
\lsb3\ell\tilde{I}_{0,2} + 15(\ell+1) \frac{\tilde{I}_{2,2}}{e^{2}}
+(-1)^{\ell+1} (2\ell+3) (2\ell+5) \frac{\tilde{I}_{2\ell+2,2}}{e^{2\ell+2}}\rsb\,,
\ea
\ba \label{eq:alpha3S}
\alpha_{n,\ell}^{(3,\bar{S})}&=&
\frac{1}{2\sqrt{1-e^2}}\lsb 3\ell\tilde{I}_{0,2} + 15\ell \frac{\tilde{I}_{2,2}}{e^{2}} - 5\delta \tilde I_{1,2} + (-1)^{\ell}(2\ell+3)(2\ell+5)
\frac{2\ell+1}{4\ell+3}
\frac{\delta\tilde{I}_{2\ell+1,2}}{e^{2\ell}}\rsb\nn\\
&+&\frac{8}{15} \bar{C}_{n,0}+ \frac{4 (-1)^{\ell}(\ell+1)(2\ell+1) \left(2e^2+\ell\right)}{(2\ell+5)(4\ell+3)e^2} \bar{C}_{n,\ell}\,,
\ea
\ba \label{eq:alpha4M}
\alpha_{n,\ell}^{(4,\bar{M})} \left(\bar{M}_2,\bar{S}_3 \right) &=& \frac{735 \left[ 3\ell e^{2}\tilde{I}_{0,2} + 15(\ell+1)\tilde{I}_{2,2} +(-1)^{\ell+1}(2\ell+3)(2\ell+5)e^{-2\ell}\tilde{I}_{2\ell+2,2}\right] }{784 e^2\sqrt{1-e^2}\bar{C}_{n,0} - 720\left(2e^2+1\right)\sqrt{1-e^2}\bar{C}_{n,1} + 735 \left(-3 e^2 \tilde{I}_{0,2}-15 \tilde{I}_{2,2} + 5e^2 \delta\tilde{I}_{1,2} + 15\delta\tilde{I}_{3,2}\right)}\nn\\
&\times&\left(\frac{\bar{S}_3}{\bar{M}_2 \bar{B}_{n,1}}-1\right)\,,
\ea
\ba \label{eq:alpha4S}
\alpha_{n,\ell}^{(4,\bar{S})}\left( \bar{M}_2, \bar{S}_3 \right)&=&
\frac{49 (-1)^{\ell} e^{-2 \ell}}{(2 \ell+5) (4 \ell+3)}
\lb 1-\frac{\bar{S}_3}{\bar{B}_{n,1}\bar{M}_2}
\rb
\left\{ -15 \left\{8 \left(e^2-1\right) \left(2 \ell^2+3 \ell+1\right) e^{2 \ell} \left(2 e^2+\ell\right) \bar{C}_{n,\ell}
\right.\right.\nn\\
&-&\left.\left.
\sqrt{1-e^2} (2 \ell+5) \left[-5 (-1)^\ell (4 \ell+3) e^{2 \ell+2} \delta\tilde{I}_{1,2} +e^2 \left(8 \ell^3+36 \ell^2+46 \ell+15\right) \delta\tilde{I}_{2 \ell+1,2}
\right.\right.\right.\nn\\
&+&\left.\left.\left.
3 (-1)^\ell \ell (4 \ell+3) e^{2 \ell} \left(e^2 \tilde{I}_{0,2}+5 \tilde{I}_{2,2}\right)\right]\right\} -16 \left(e^2-1\right) (-1)^\ell \left(8 \ell^2+26 \ell+15\right) e^{2 \ell+2} \bar{C}_{n,0}\right\} \nn\\
&\times&\left\{784 e^2 \left(e^2-1\right) \bar{C}_{n,0}-15 \left[48 \left(2 e^4-e^2-1\right) \bar{C}_{n,1}-49 \sqrt{1-e^2} \left(15 \delta\tilde{I}_{3,2}+5 e^2 \delta\tilde{I}_{1,2}\right.\right.\right.\nn\\ &-&\left.\left.\left. 3 e^2 \tilde{I}_{0,2}-15 \tilde{I}_{2,2}\right)\right]\right\}^{-1}\,.
\ea

\ew

\bibliography{master}

\end{document}